# Reconfigurable Multi-State Optical Systems Enabled by $VO_2$ Phase Transitions

Xiaoyang Duan,[1] Samuel T. White,[2] Yuanyuan Cui,[3] Frank Neubrech,[1] Yanfeng Gao,[3,*] Richard F. Haglund,[2,4,*] Na Liu[5,6,*]

[1]Max Planck Institute for Intelligent Systems, Heisenbergstrasse 3, 70569 Stuttgart, Germany
[2]Department of Physics and Astronomy, Vanderbilt University, Nashville, TN 37212, USA.
[3]School of Materials Science and Engineering, Shanghai University, Shanghai 200444, China.
[4]Interdisciplinary Materials Science Program, Vanderbilt University, Nashville, TN 37212, USA.
[5]2nd Physics Institute, University of Stuttgart, Pfaffenwaldring 57, 70569 Stuttgart, Germany
[6]Max Planck Institute for Solid State Research, Heisenbergstrasse 1, 70569 Stuttgart, Germany
*E-mail: na.liu@pi2.uni-stuttgart.de, richard.haglund@vanderbilt.edu, yfgao@shu.edu.cn.

**ABSTRACT:** Reconfigurable optical systems are the object of continuing, intensive research activities, as they hold great promise for realizing a new generation of compact, miniaturized and flexible optical devices. However, current reconfigurable systems often tune only a single state variable triggered by an external stimulus, thus leaving out many potential applications. Here we demonstrate a reconfigurable multi-state optical system enabled by phase transitions in vanadium oxide ($VO_2$). By controlling the phase-transition characteristics of $VO_2$ with simultaneous stimuli, the responses of the optical system can be reconfigured among multiple states. In particular, we show a quadruple-state dynamic plasmonic display that responds to both temperature tuning and hydrogen-doping. Furthermore, we introduce an electron-doping scheme to locally control the phase-transition behaviour of $VO_2$, enabling an optical encryption device encoded by multiple keys. Our work points the way toward advanced multi-state reconfigurable optical systems, which substantially outperform current optical devices in both breadth of capabilities and functionalities.

Developing functional optical systems that can respond to multiple external stimuli is one of the central objectives in photonics. In general, there are two routes to accomplishing this goal. One is to create active optical elements directly,[1-11] so that the resulting plasmonic,[3-7] dielectric,[8] or cavity resonators,\[9-11] exhibit tunable degrees of freedom. At visible frequencies, the choices of active materials, however, are not plentiful. Endowing plasmonic resonators, in particular, with tunability or reconfigurability is especially challenging, because they are often made of noble metals that are unresponsive to chemical tuning. The alternate route is to utilize active materials as surrounding media for optical resonators.[12-17] In this case, the material library becomes expansive. Among the available active materials, vanadium dioxide ($VO_2$) is particularly attractive, given the remarkable range of electronic and structural properties accessible to thermal, electrical and optical excitation and to stress.[18-21] As a correlated material, $VO_2$ undergoes a reversible insulator (monoclinic, m-$VO_2$)-to-metal (rutile, r-$VO_2$) phase transition (IMT) through temperature tuning, resulting in drastic changes in electronic and optical properties.[18-21] In addition, $VO_2$ is very sensitive to extrinsic doping, for instance, by hydrogen (H), lithium (Li) and other transition metals.[22-28] Recently, it has been demonstrated that light hydrogenation of $VO_2$ can result in a metallic phase, H-$VO_2$(M) at room temperature, while heavy hydrogenation gives rise to an insulating phase, H-$VO_2$(I).[22-24] This panoply of distinct $VO_2$ states thus make it an ideal reconfigurable material to incorporate into photonic devices.

However, so far only temperature tuning of $VO_2$ has been widely exploited for photonic applications,[14, 15, 29-35] leaving out many potential applications that could be enabled by $VO_2$. Here we demonstrate multi-state optical systems reconfigured by $VO_2$ phase transitions, which can be initiated by thermal tuning, H-doping, and electron-doping (*e*-doping). We first show a quadruple-state dynamic plasmonic display that responds to both temperature tuning and H-doping. We also exhibit a powerful scheme to locally control the phase-transition behaviour of $VO_2$ by *e*-doping, instantiated by metals deposited on the VO$_2$ surface. At the VO$_2$/metal interface, electrons can be injected from a metal with lower work function to VO$_2$ or from VO$_2$ to a metal with higher work function, leading to facile alteration of the VO$_2$ phase-transition behaviour. Finally, we demonstrate an optical encryption device with two-level information encoded by multiple keys, including temperature, *e*-doping, and H-doping.

Our work provides profound new insights into the phase-transition mechanisms as well as novel examples of the versatile optical functionality of VO$_2$. Moreover, it advances the perspectives of technologically important research by combining the strength of optics and functional materials to enable new, real-world applications.

**RESULTS AND DISCUSSION**

**Hydrogen-doping of VO$_2$**

Vanadium dioxide (60 nm) films are deposited on gold (Au, 100 nm)/silicon (Si) substrates by radio-frequency (RF) magnetron sputtering (see Experimental Section, Supporting Information). The thick Au film employed here as reflective backplane for the optical devices that will be discussed later. Equally important, the polycrystalline VO$_2$ films on Au can sustain long hydrogen exposure without any damage, but not when deposited on a Si substrate (see Figure S1, Supporting Information). On gold, the deposited VO$_2$ film grains tend to have a preferred growth orientation with the monoclinic (011)$_m$ plane parallel to the substrate surface, as revealed by X-ray diffraction (XRD) measurements (Figure S2, Supporting Information). Nano-sized Pd dots, which can efficiently catalyse the dissociation of hydrogen molecules into atoms, are deposited on top of the VO$_2$ films to facilitate hydrogenation in the temperature range of interest between 20 °C and 100 °C. A schematic and scanning electron microscope (SEM) image of the sample are shown in Figure 1a and b, respectively. The phase-transition paths of VO$_2$ are investigated by *in situ* optical reflectance spectroscopy at different H-doping levels and temperatures as shown in Figure 1c. The reflectance intensity at a fixed wavelength of 600 nm ($R_{600}$) is tracked during the different doping procedures (see Figure S3, Supporting Information). To identify and further confirm the formation of different states of $VO_2$, electrical resistance across the sample is also measured as a function of temperature and presented simultaneously with the optical measurements in Figure 1d (Details in Figure S4, Supporting Information).

As shown in Figure 1c, unhydrogenated $VO_2$ undergoes the IMT between the m-$VO_2$ phase and the r-$VO_2$ phase by temperature tuning in approximately 15 min, when heating and cooling rates of 0.1°C are used. The heating and cooling paths are indicated by '①→②→③' and '③→④→①' (black curves), respectively. The phase-transition temperature is defined as $T_c = (T_{c,\uparrow} + T_{c,\downarrow})/2$, where $T_{c,\uparrow}$ corresponds to the m- to r-$VO_2$ transition temperature, and $T_{c,\downarrow}$ is the reverse r- to m-$VO_2$ transition temperature; the measured $T_c$ of this sample is 64 °C. During hydrogenation ($H_2$, 20 vol.%), two new phases are observed that are much less dependent on temperature variations. One is the lightly-doped H-$VO_2$(M) phase (green curve) and the other is the heavily-doped H-$VO_2$(I) phase (pink curve). The *in situ* resistance measurements at different temperatures and H-doping levels (Figure 1d) also reveal the four states, m-$VO_2$, r-$VO_2$, H-$VO_2$(M), and H-$VO_2$(I), consistent with the optical measurements in Figure 1c.

To understand the relation between the H-doping level and temperature, hydrogenation ($H_2$, vol. 20%) and dehydrogenation ($O_2$, vol. 20%) in different states and at various temperatures are investigated. Four representative initial states are selected, *i.e.*, ① m-$VO_2$ at 40 °C, ② (m/r-$VO_2$)$_\uparrow$ at 60 °C, ③ r-$VO_2$ at 100 °C, and ④ (m/r-$VO_2$)$_\downarrow$ at 60 °C as marked in Figure 1c. Due to hydrogenation, ①, ②, ③ and ④ all first reach the H-$VO_2$(M) phase and then the H-$VO_2$(I) phase. The switching times of light hydrogenation at 40 °C, 60 °C, and 100 °C are ~30 h, ~6 h, and ~30 min accordingly. The switching time of heavy hydrogenation at 100 °C is ~14 h. In the case of dehydrogenation, when starting from the H-VO$_2$(M) phase, at 40 °C and 100 °C, ① and ③ can be restored, respectively, whereas at 60 °C that is close to $T_c$, dehydrogenation only leads to ④, regardless of whether the initial is being ② or ④. The switching times of dehydrogenation at 40 °C, 60 °C, and 100 °C are ~180 h, ~24 h and ~1 h accordingly. More interestingly, when starting from the H-VO$_2$(I) phase, dehydrogenation follows the same aforementioned paths for the respective cases, completely bypassing the H-VO$_2$(M) phase. (The details of hydrogenation and dehydrogenation dynamics can be found in Supporting Information.) Our observations by optical spectroscopy agree well with the work reported by Yoon *et al.* using resistivity measurements.[23] Figure 1e summarizes the relationship of the phase-transition trajectories to temperature and H-doping level. All the corresponding reflectance spectra correlated to the H-doping can be found in Figures S5 to S9, Supporting Information.

**Quadruple-state dynamic plasmonic display**

The wide range of phase-transition states of VO$_2$ and the accompanying drastic changes in optical properties form a solid basis to realize dynamic optical devices with diverse responses. In the following, we demonstrate a quadruple-state dynamic plasmonic display by capitalizing on the tunability of VO$_2$ in response to a combination of temperature and H-doping. Figure 2a shows the schematic of the plasmonic display. Stacked aluminum (Al)/Al$_2$O$_3$ nanodisks reside on a substrate comprising Pd dots on a VO$_2$ (60 nm)/Au mirror substrate; the Al and Al$_2$O$_3$ nanodisks are both 30 nm tall. In order to generate a palette of colours with a wide range, the diameter ($D$) and interparticle gap ($g$) of the Al/Al$_2$O$_3$ nanodisks are tuned stepwise, respectively. Figure 2b shows the SEM image of a representative area ($D$ = 102 nm, $g$ = 100 nm) of the palette; a tilted view is presented as inset in Figure 2b. The optical microscopy images of the palette at the four different states of VO$_2$ [m-VO$_2$, r-VO$_2$, H-VO2(M), and H-VO$_2$(I)] are shown in Figure 2c. Notable colour changes are visualized on the palette in these four states.

To understand the observed colour changes, experimental and simulated reflectance spectra of the selected colour squares (i–iv in Figure 2c) are presented using solid and dashed lines in Figure 2d, respectively. The simulated spectra agree well with the experimental results. For each state, the resonance (*i.e.*, reflectance dip) exhibits a clear red shift when *D* increases at a fixed *g*; for instance, see the resonance positions marked by the blue arrows for the cases of the m-VO$_2$ state in Figure 2d. To provide further insights, surface charge and magnetic field distributions at resonance for colour square *iii* at all four states are shown in Figure S10, Supporting Information, and Figure 2e, respectively. At each state, the resonance variations result from the refractive index changes of VO$_2$. However, the resonances are more pronounced at the m-VO$_2$ (blue) and H-VO$_2$(I) (pink) states than at the r-VO$_2$ (red) and H-VO$_2$(M) (green) states, because m-VO$_2$ and H-VO$_2$(I) are insulators and function as dielectric spacers along with the Al$_2$O$_3$. The Al particle plasmons strongly interact with the bottom Au mirror with the electromagnetic fields localized in Al$_2$O$_3$, as in a typical particle-on-mirror configuration.[36] On the other side, r-VO$_2$ and H-VO$_2$(M) are bad metals, when compared to Au. The interactions between the Al particle plasmons and the bad-metal substrates broaden and smear the resonances, an effect exacerbated by lower reflectance spectral profiles. In other words, the alterations to the optical spectra, in terms of resonance position and linewidth as well as reflected intensity give rise to the distinctive colour changes at the four states as shown in Figure 2c. To quantify the colour tunability, the chromaticity coordinates of the palette are calculated for the four different states using the colour-matching functions defined by the International Commission on Illumination (CIE)[37] and shown in Figure S11, Supporting Information.

Such abundant colours and distinct colour transformations enable the realization of arbitrary microprints with multiple dynamic states. As a demonstration, a plasmonic colour display with resolution as high as 120,000 dpi is implemented using Vincent van Gogh's *The Starry Night* as design blueprint. The optical microscope images of the dynamic plasmonic display in the four different states are presented in Figure 2f. The accompanying movie recording the colour transformations can be found in Movie S1, Supporting Information. Notably, there is no performance degradation over 100 cycles, even after the sample has been placed in ambient air for one year, demonstrating the excellent reversibility and durability of the device.

**Electron-doping of VO$_2$**

To further demonstrate the versatile controllability of this material system, a local *e*-doping mechanism is used to modulate the phase-transition behaviour of VO$_2$. Ultra-thin metal layers (~1.5 nm) are deposited on a VO$_2$ (60 nm)/Au (100 nm)/Si substrate at different locations, as shown in the optical image (Figure 3a). To evaluate the effects resulting from the surface metallization with high accuracy, the reflectivity of the areas covered by different metal species are measured simultaneously (see the blue-dashed frame in Figure 3a) over the temperature range from 20 °C to 90 °C.

Figure 3b shows the hysteresis curves for VO$_2$ covered by different ultra-thin metals. For better comparisons, the hysteresis curves are normalized, spanning the range from 0 to 1. The deduced phase-transition temperatures are summarized in Figure 3c and the $T_c$ of VO$_2$ without surface metallization is indicated by the black-dashed line in the same plot. It is evident that the surface metallization of VO$_2$ has a substantial influence on the phase-transition behaviour. More specifically, tuning of the transition temperatures is clearly observed, following the order from high to low temperatures, $T_{c,Pt} > T_{c,Au} > T_{c,VO_2} > T_{c,Pd} > T_{c,Cu} > T_{c,Ni} > T_{c,Ti} > T_{c,Al} > T_{c,Cr}$. The experimental results also show that the surface metals affect the phase-transition behaviour more drastically along the heating path than the cooling path (see Figure 3b).

In order to elucidate the underlying physics, Figure 4a presents the work function values of the different metals,[38, 39] which exhibit the order $W_{Pt} > W_{Au} \gtrsim W_{VO_2} \gtrsim W_{Pd} > W_{Cu} > W_{Ni} > W_{Ti} > W_{Al} > W_{Cr}$ and reveal a clear correlation between the phase-transition temperature of $VO_2$ and the work function of the deposited metal.

A pure electron-doping (hereinafter, $e$-doping) mechanism is proposed to interpret the experimental observations. Due to the work-function difference, metals (*i.e.*, Cr, Al, Ti, Ni, Cu) with lower work-function values and thus higher Fermi levels can donate electrons across their interface to the $VO_2$, until the Fermi levels are aligned, leading to increased carrier concentration in $VO_2$. According to the Mott mechanism, the IMT of $VO_2$ is driven by an increase of electron concentration in the conduction band.[40, 41] Thus, once the electron concentration in m-$VO_2$ reaches a critical value, the IMT is triggered and the metallic r-$VO_2$ phase begins to nucleate. More specifically, the increased carrier concentration resulting from the surface metal contact, for instance, Cr, lowers the energy barrier across the IMT and destabilizes the m-$VO_2$ phase, thus effectively decreasing the transition temperature. In contrast, for metals (*i.e.*, Pt), which have higher work-function values and thus lower Fermi levels, electrons are donated from $VO_2$ to the metals at their interfaces, leading to an increased transition temperature. The schematics in Figure 4b depict the two exemplary cases, in which electrons flow from Cr to $VO_2$ (from $VO_2$ to Pt) at the $VO_2$/Cr ($VO_2$/Pt) interfaces, respectively.

This interpretation is supported by the first-principle calculations. Figure 4c presents the calculated differential charge distributions, which reveal electron injection from Cr to $VO_2$ and from $VO_2$ to Pt at the $VO_2$/Cr and $VO_2$/Pt interfaces, respectively. With a local metal contact, the atoms in the surface layer of $VO_2$ are reconstructed due to the $VO_2$/metal lattice mismatch.

However, the lattice parameters of the atomic layers beneath the VO$_2$ surface are very close to those of pure VO$_2$. For instance, the relative lattice mismatch beneath the VO$_2$/Cr and VO$_2$/Pt interfaces is below 0.07% and 0.04%, respectively. Such lattice mismatches resulting from $e$-doping are significantly smaller than those introduced by H-doping (>1%). In this regard, electron loading/unloading in VO$_2$ can be readily achieved by introducing surface metals on VO$_2$ with negligible lattice distortions.

The effects of the surface metals are also manifested in changes of the electronic structure of VO$_2$ as shown in Figure 4d. The O-2$p$ and V-3$d$ electrons contribute significantly to the valence band ($E_v$) and conduction band ($E_c$), respectively, where $E_c - E_v$ defines the band gap energy $E_{gap}$. The position of the Fermi level in the band gap, namely, $E_F - E_v$, indicates the relative electron concentration in VO$_2$. By comparing VO$_2$ with and without surface metal, it is apparent that electrons are donated from Cr to VO$_2$ in the case of VO$_2$/Cr, whereas VO$_2$ donates electrons to Pt in the case of VO$_2$/Pt. Unlike from H-doping, $e$-doping spatially localizes the IMT of VO$_2$ along both directions, that is, the phase-transition temperature can be increased or decreased simply by attaching an ultra-thin surface metal with high or low work function on the surface, respectively.

**Two-level optical encryption**

The ability to regulate the phase-transition behaviour of VO$_2$ by multiple means — including temperature, H-doping, and $e$-doping — offers a novel route to realizing high-level optical encryption systems protected by multiple keys. To this end, we further demonstrate a dynamic display encrypted with two-level optical information. Figure 5a shows the design schematic of the display, which contains two merged patterns. One is a QR code that links to our website 'www.is.mpg.de/liu' and the other is a word 'VO$_2$'. There are four functional zones: The white area is VO$_2$ and the blue area is VO$_2$/Pd (1 nm), both exhibiting the IMT (see black curves) at approximately 71 °C along the heating path. The Pd, needed to "crack" adsorbed hydrogen, exerts negligible influence on the IMT behaviour, as shown in Figure 3b. In the temperature range of interest between 30 °C and 100 °C, the former is unreactive to $H_2$, whereas the latter reacts to $H_2$. The grey area is $VO_2$/Ti (2 nm) that is unreactive to $H_2$ while the red area is $VO_2$/Ti (2 nm)/Pd (1 nm) that is reactive to $H_2$. Both areas undergo the IMT (see grey curves) at approximately 64 °C along the heating path.

At the initial state (m-$VO_2$) as shown in Figure 5c, the display is a yellowish colour with blank information (state *i*). To read out the QR code pattern, temperature is utilized as the first decryption key. With increasing temperature, the areas covered by Ti (grey and red areas in Figure 5a) follow the IMT path shown in the upper grey curve. These areas start to become greenish in the temperature range of 64 ± 3 °C. The highest optical contrast of the QR code pattern is achieved at 64 °C, reaching state *ii*. With further temperature increase above 80 °C, the QR code disappears and the display shows greenish colour with blank information (state *iii*), as all the four areas are transformed to r-$VO_2$. The temperature-controlled process is fully reversible. States *i* and *ii* can be subsequently restored by decreasing the temperature. To read out the '$VO_2$' pattern, hydrogen is employed as the second decryption key. Upon loading $H_2$, the areas with Pd coverage (blue and red areas in Figure 5a) transition to H-$VO_2$(M) and the pattern '$VO_2$' becomes clearly visible (state *iv*). The hydrogen-regulated process is also fully reversible, as state *i* can be restored by loading $O_2$.

**CONCLUSIONS**

In summary, we have demonstrated a $VO_2$-based multi-state optical systems for dynamic optical display applications. The phase-transition behavior and the optical functions of $VO_2$ are modulated by temperature, H-doping, and *e*-doping. Temperature tuning leads to the phase transition between r-VO$_2$ and m-VO$_2$, while H-doping introduces two additional hydrogenated phases, H-VO$_2$(M) and H-VO$_2$(I). Electron-doping provides a facile method to control the phase-transition behavior by utilizing surface metal contacts. Finally, by depositing ultra-thin metal coupons with different work functions, the transition temperature can be readily tuned, as a result of the electron-transfer process at the VO$_2$/metal interface that aligns the Fermi levels.

Further studies combining unique material functionalities, smart optical designs and other pairs of complementary, external excitations that drive phase transitions (*e.g.,* light and pressure in VO$_2$) can stimulate new methodological perspectives in both materials science and optics. In turn, these studies may shed light on a new generation of multi-level functional optical devices for applications in optical information storage, optical encryption with high security levels, and high-resolution optical and holographic display technologies.

## ASSOCIATED CONTENT
The authors declare no competing financial interest

## AUTHOR INFORMATION
**Corresponding Author**
*E-mail: na.liu@pi2.uni-stuttgart.de, richard.haglund@vanderbilt.edu, yfgao@shu.edu.cn.

**ORCID**
Na Liu: 0000-0001-5831-3382
Richard Haglund: 0000-0002-2701-1768
Xiaoyang Duan: 0000-0002-8720-3788
Samuel White: 0000-0002-3168-8894
Frank Neubrech: 0000-0002-5437-4118
Yuanyuan Cui: 0000-0003-2971-0543
Yanfeng Gao: 0000-0001-7751-1974

**Author Contributions**
X.D. and N.L. conceived the project. X.D. designed and performed the experiments, and analyzed the data. S.T.W. prepared and characterized the VO$_2$ films. Y.C. conducted the first-principles calculations. X.D. carried out the fitting of dielectric constants of $VO_2$ and the optical simulations. All authors discussed the results and commented on the manuscript.

**Acknowledgements**
We gratefully acknowledge the generous support by the 4th Physics Institute at the University of Stuttgart for the usage of clean room facilities. We also thank Helga Hoier for helping with the XRD measurements, and Ksenia Rabinovich for the ellipsometry measurements. This project was supported by the European Research Council (ERC *Dynamic Nano*) grant, the Max Planck Society (Max Planck Fellow), and by a contract with Triton Systems, Inc. to Vanderbilt University through a Small Business Innovative Research program of the National Aeronautics and Space Agency (S. T. W.). Y.G. and Y.C. gratefully acknowledge the support from the National Natural Science Foundation of China (51873102 and 51972206).

**Supporting Information**
Sample fabrications, hydrogenation/dehydrogenation, optical characterizations, numerical simulations, first-principle calculations, and additional experimental data, Pages S1–S10, Figures S1–S12, Movie S1. This material is available free of charge via the internet at http://pubs.acs.org.

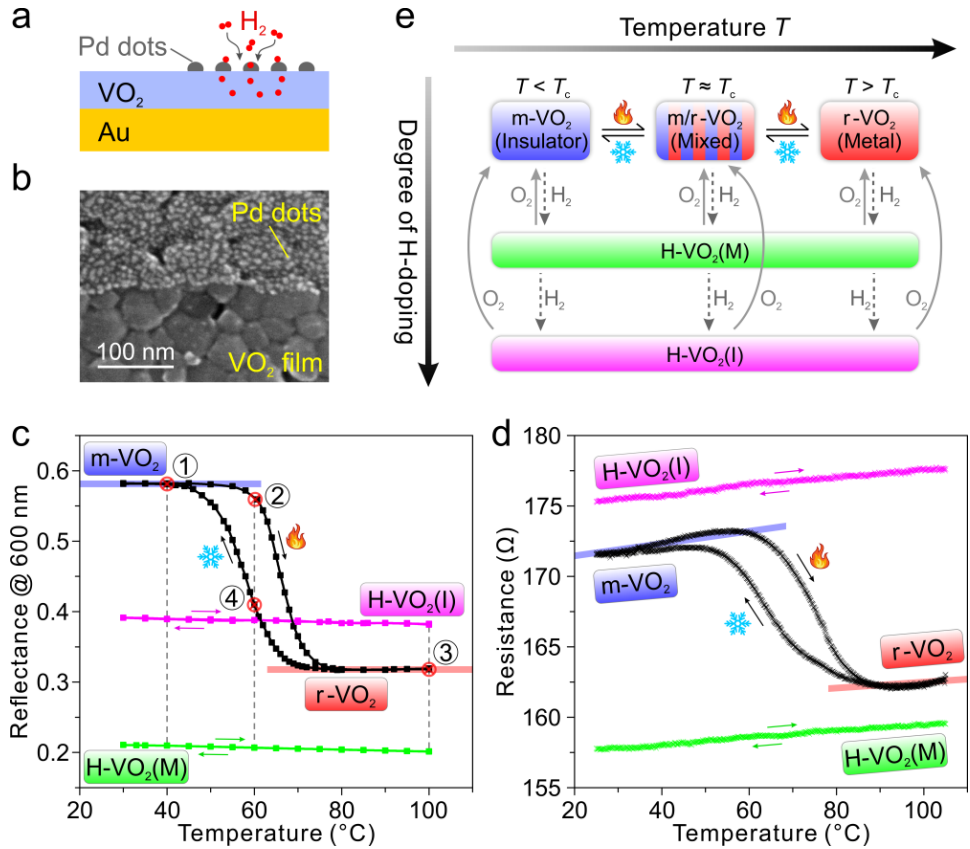

**Figure 1. Tuning the phase transition behaviour of VO₂ by temperature and H-doping.** a) Schematic illustration of the sample configuration. b) SEM image of the sample. Nano-sized Pd dots on the VO₂ surface enable hydrogenation and dehydrogenation at ambient temperature. c) Reflectance of the sample at 600 nm ($R_{600}$) measured at different temperatures and H-doping levels. VO₂ exhibits a reversible IMT between the insulating m-VO₂ phase and the metallic r-VO₂ phase. '①→②→③' and '③→④→①' indicate the heating and cooling paths (black curves), respectively. H-doping introduces two hydrogenated phases, H-VO₂(M) (green curve) and H-VO₂(I) (pink curve). d) Resistance measurements of the sample carried out simultaneously with the optical measurements. e) Summary of phase-transition paths deduced from optical measurements. Mixed phase (m/r-VO₂) appears at temperatures close to the phase-transition temperature $T_c$.

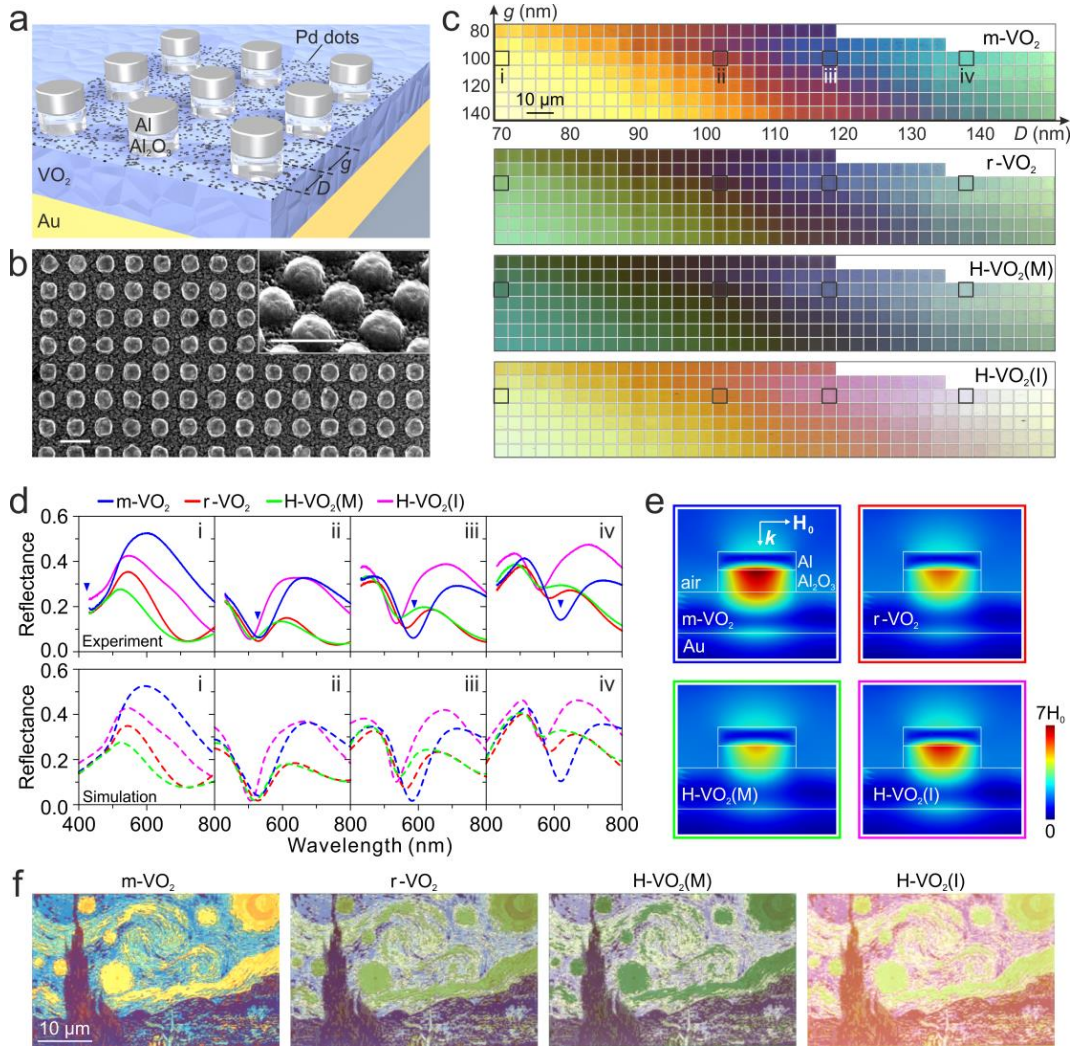

**Figure 2. Quadruple-state dynamic color display.** a) Schematic of the quadruple-state dynamic color display. Al$_2$O$_3$/Al nanodisks with different diameter ($D$) and interparticle gap ($g$) reside on a VO$_2$ (60 nm)/Au mirror substrate. Nano-sized Pd dots on VO$_2$ are utilized to facilitate its hydrogenation and dehydrogenation. b) Overview SEM image of a palette square with $D$ = 102 nm and $g$ = 100 nm. Inset: enlarged tilted view of the SEM image. Scale bar: 200 nm. c) Optical micrographs of a colour palette with stepwise tuning of $D$ and $g$ at the m-VO$_2$, r-VO$_2$, H-VO$_2$(M), and H-VO$_2$(I) states accordingly. Representative color squares *i–iv* for spectral characterizations are highlighted. d) Experimental and simulated reflectance spectra of the selected colour squares from (c) at the four different states. Blue arrows indicate the resonance dip at the m-VO$_2$ state. e) Simulated magnetic field distributions at resonance of the color square *iii* ($D$ =118 nm and $g$ = 100 nm) at the four different states. f) Optical micrographs of the dynamic color display of Vincent van Gogh's *The Starry Night* at the four different states, showing vivid colour tuning. (Gogh, Vincent van (1853-1890): *The Starry Night*, 1889. New York, Digitale (1) Museum of Modern Art (MoMA). Oil on canvas, 29 × 36 OE (73.7 × 92.1 cm). Acquired through the Lillie P. Bliss Bequest. Acc. no.: 472.1941. © 2017. Digital image, The Museum of Modern Art, New York/Scala, Florence.)

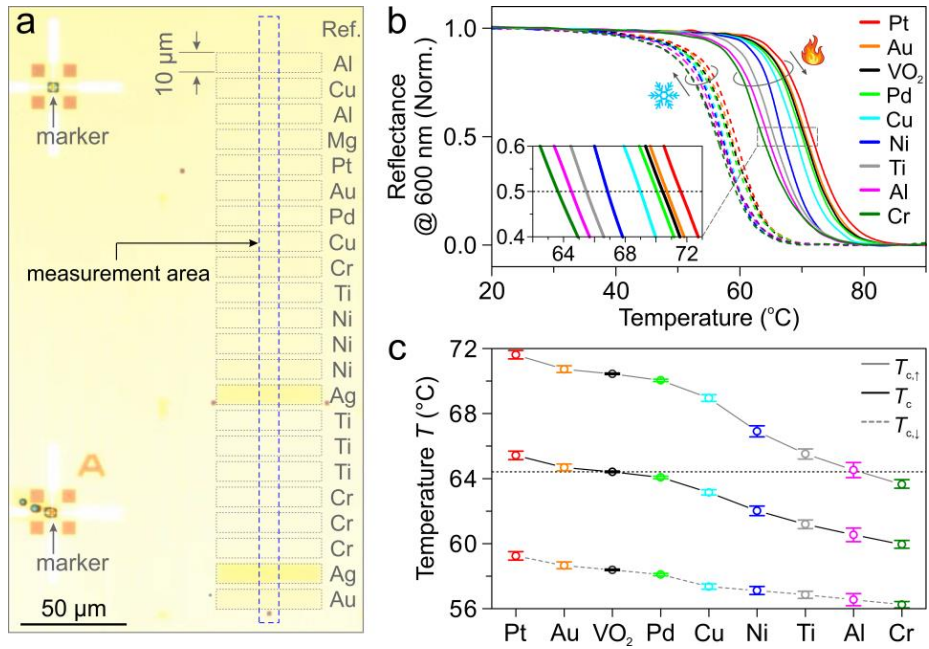

**Figure 3. Tuning the phase transition behaviour of VO$_2$ by local *e*-doping.** a) Optical image of the sample layout. Different metals (Pt, Au, Pd, Cu, Ni, Ti, Al, Cr, etc.) are deposited on a VO$_2$ (60 nm)/Au mirror surface. The optical spectra of the VO$_2$ covered by different metals (blue-dashed frame, 10 μm width) are measured simultaneously. b) Normalized hysteresis curves of the VO$_2$ covered by different metals (~1.5 nm thick each). Heating and cooling paths are presented using solid and dashed lines, respectively. c) Transition temperatures $T_{c,\uparrow}$, $T_{c,\downarrow}$, and $T_c$ deduced from (b). Gray-dotted line indicates the transition temperature of VO$_2$ without any surface metal.

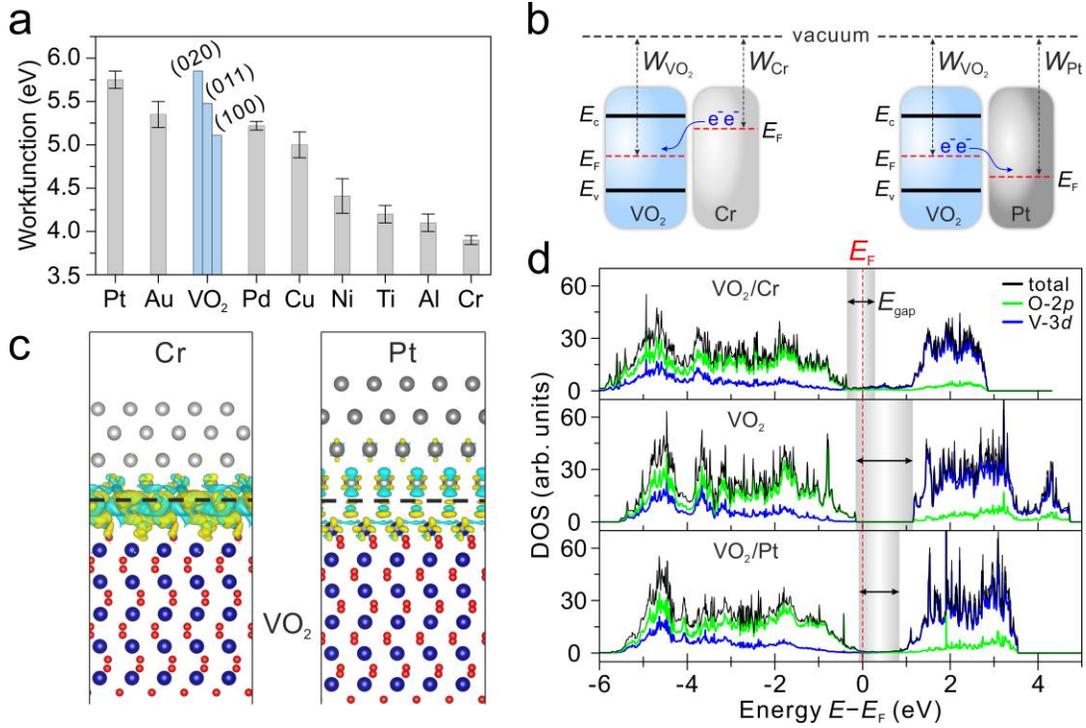

**Figure 4. Relation between the phase-transition temperature of $VO_2$ and the surface metal work function.** a) Experimental work function values for different metals and calculated work function values for $VO_2$. b) Schematic depiction of electrons flowing from a metal with a lower work function value (*e.g.* Cr) to $VO_2$ (left panel), and flowing from $VO_2$ to a metal with higher work function value (*e.g.* Pt, right panel). $E_F$: Fermi level. c) Calculated differential charge distributions at the $VO_2$/Cr (left panel) and $VO_2$/Pt (right panel) interfaces, respectively. Cyan and yellow bubbles represent hole and electron charges, respectively. Blue, red, gray, and dark gray balls represent V, O, Cr, and Pt atoms. d) Density of states (DOS) of the $VO_2$/Cr, pure $VO_2$, and $VO_2$/Pt systems. $E_{gap}$ in each case is highlighted in gray.

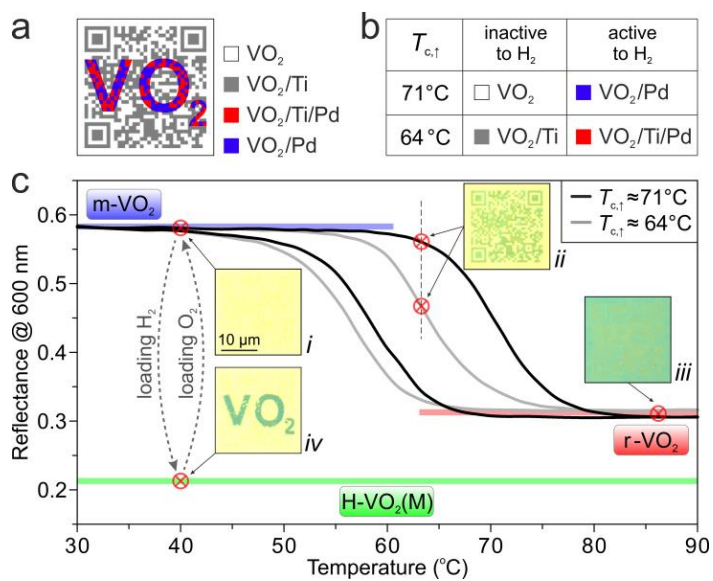

**Figure 5. Dual-key information encryption.** a) Design of a dynamic optical display with two-fold information encryption. The two patterns are a QR code and a word 'VO$_2$', respectively. White area: VO$_2$, Gray area: VO$_2$/Ti (2 nm), Red area: VO$_2$/Ti (2 nm)/Pd (1 nm), and blue area: VO$_2$/Pd (1 nm). b) Phase-transition temperatures along the heating path and response to H$_2$ for different areas. c) Four states (*i*, *ii*, *iii*, and *iv*) of the optical display. The two decryption keys are temperature and hydrogenation, respectively.

For Table of Contents Use Only

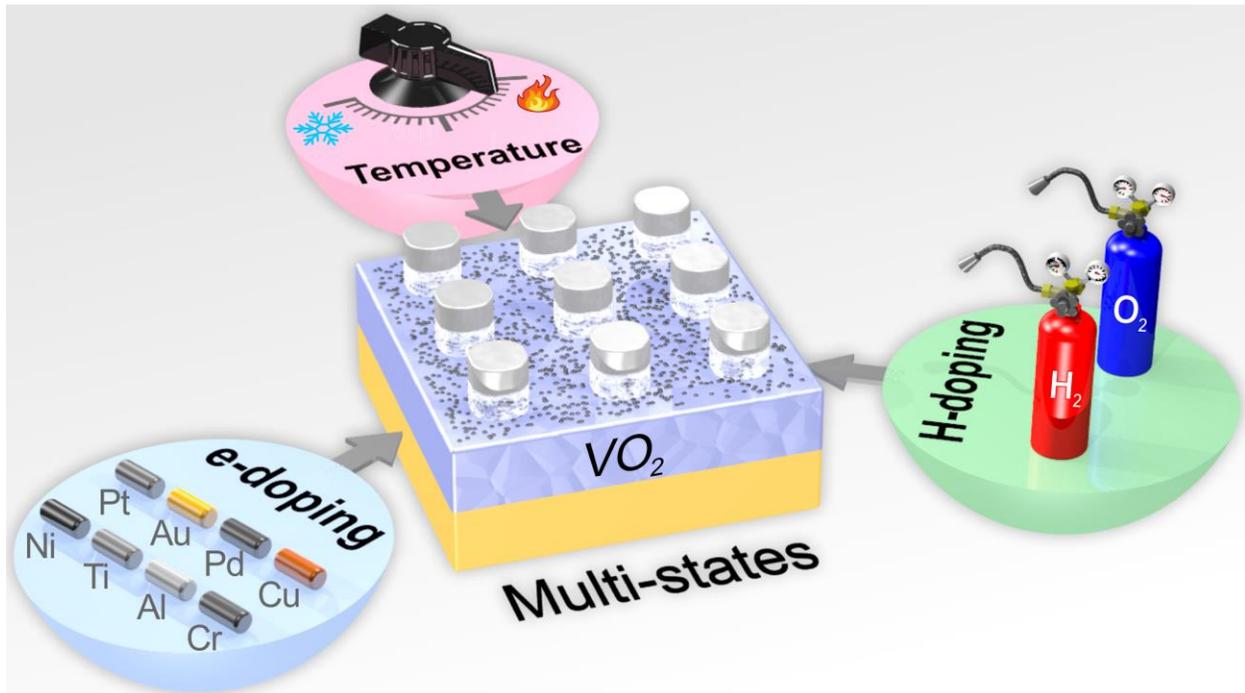

Title: Reconfigurable Multi-State Optical Systems Enabled by $VO_2$ Phase Transitions

Authors: Xiaoyang Duan, Samuel T. White, Yuanyuan Cui, Frank Neubrech, Yanfeng Gao, Richard F. Haglund, Na Liu

Brief synopsis: The authors demonstrate a reconfigurable multi-state optical system enabled by the phase transitions of $VO_2$, which can be triggered by multiple stimuli, including temperature tuning, hydrogen doping, and electron doping.